\documentclass[referee]{aa} 

\begin{document}
\def\al{&\!\!\!\!}
\def\x{{\bf x}}
\def\f{\frac}
   \thesaurus{03      A\&A Section 3: dark matter, missing mass,
   alternative dynamics, $f(R)$ gravitation
               }  

   \title{An $ f(R)$ gravitation for galactic environments}

   \author{Y. Sobouti}

      \offprints{Y. Sobouti}
      \institute{Institute for Advanced Studies in Basic Sciences, P. O. B. 45195-1159, Zanjan,
      Iran\\
      email:  sobouti@iasbs.ac.ir}

    \date{14, March 2006 }
\titlerunning{An $ f(R) $ gravitation for galactic environments}

\maketitle

   \begin{abstract}
We propose an action-based $ f(R) $ modification of Einstein's
gravity that admits a modified Schwarzschild-deSitter metric. In
the weak field limit this amounts to adding a small logarithmic
correction to the newtonian potential. A test star moving in such
a spacetime acquires a constant asymptotic speed at large
distances. This speed, calibrated empirically, is proportional to
the fourth root of the mass of the central body in compliance with
the Tully-Fisher relation. A variance of MOND's gravity emerges as
an inevitable consequence of the proposed formalism. It has also
been shown (Mendoza et al. 2006) that a) the gravitational waves
in this spacetime propagate with the speed of light in vacuum and
b) there is a lensing effect added to what one finds in the
classic GR.
   \end{abstract}

\section{Introduction}

Convinced of cosmic speed up and not finding the dark energy
hypotheses a compelling explanation, some cosmologists have looked
for alternatives to Einstein's gravitation (Deffayet et al. 2002;
Freese et al. 2002; Ahmed et al. 2002; Dvali et al. 2003;
Capozziello et al. 2003; Carroll et al. 2003; Norjiri et al. 2003,
2004, and 2006; Das et al. 2005; Sotiriou 2005; Woodard 2006).
There is a parallel situation in galactic studies. Dark matter
hypotheses, intended to explain the flat rotation curves of
spirals or the large velocity dispersions in ellipticals, have
raised more questions than answers.

Alternatives to newtonian dynamics have been proposed but have had
their own critics. Foremost among such theories, the Modified
Newtonian Dynamics (MOND) of Milgrom (1983 a,b,c) is able to
explain the flat rotation curves (Sandres et al. 1998 and 2002)
and justify the Tully-Fisher relation with considerable success.
But it is often criticized for the lack of an axiomatic
foundation; see, however, Bekenstein's (2004) TeVeS theory where
he attempts to provide such a foundation by introducing a tensor,
a vector, and a scalar field into the field equations of GR.

Here we are concerned with galactic problems. We suggest following
cosmologists and look for a modified Einstein gravity tailored to
galactic environments. In Sects. 2 and 3 we design an action
integral, different but close to that of Einstein-Hilbert, and
find a spherically symmetric static solution to it. In Sect. 4 we
analyze the orbits of test objects moving in this modified
spacetime and demonstrate the kinship of the obtained dynamics
with MOND. Section 5 is devoted to concluding remarks.\textbf{}


\section{A modified field equation}
The model we consider is an isolated mass point. As an alternative
to the Einstein-Hilbert action, we assume
\begin{eqnarray} \label{e1}
&&S= \frac{1}{2}\int f(R)\sqrt{-g}d^4x,
\end{eqnarray}
where $R$ is the Ricci scalar and $f(R)$  an as yet unspecified,
but differentiable function of $R$. Variations in $S$ with respect
to the metric tensor lead to the following field equation
(Capozziello et al. 2003):
\begin{eqnarray} \label{e2}
&&R_{\mu\nu}-\frac{1}{2}g_{\mu\nu}\frac{f}{h}=\left(h_{;\mu\nu}-{h_{;_\lambda}}^\lambda
g_{\mu\nu}\right)\frac{1}{h},
\end{eqnarray}
where $h=df/dR$. The case $f(R)=R+constant$ and $h=1$ gives the
Einstein field equation with a cosmological constant included in
it. For the purpose of galactic studies, we envisage a spherically
symmetric static Schwarzschild-like metric,
\begin{eqnarray}\label{e3}
&&ds^2=-B(r)d t^2+ A(r) d r^2+ r^2\left(d\theta^2+\sin^2\theta
d\varphi^2\right).
\end{eqnarray}\vspace{1mm}
From Eqs. (\ref{e2}) and (\ref{e3}) one obtains
\begin{eqnarray}\label{e4}
&& \frac{B'}{B}+\frac{A'}{A}=-r\frac{h''}{h}+\frac{1}{2}
r\left(\frac{B'}{B}+\frac{A'}{A}\right)\frac{h'}{h},
\\&&\cr \label{e5}
&&\frac{B''}{B}-\frac{1}{2}\left(\frac{B'}{B}+\frac{2}{r}\right)\left(\frac{B'}{B}
+\frac{A'}{A}\right)-\frac{2}{r^2}+\frac{2A}{r^2}\cr
&& \hspace{6mm}= 2\frac{h''}{h}-(\frac{A'}{A}+\frac{2}{r})\frac{h'}{h} ,\\
&& \cr\label{e6}\al\al
\frac{B''}{B}-\frac{1}{2}\frac{B'}{B}\left(\frac{B'}{B}+\frac{A'}{A}\right)-
\frac{2}{r}\frac{A'}{A}\cr \al\al \hspace{6mm}= f\frac{A}{h}
-\left(\frac{B'}{B}+\frac{4}{r}\right)\frac{h'}{h},\\
&&\cr \label{e7} &&
R=2\frac{f}{h}-\frac{3}{A}\left[\frac{h''}{h}+\left\{\frac{1}{2}\left(
\frac{B'}{B}-\frac{A'}{A}\right)+\frac{2}{h}\right\}\frac{h'}{h}\right].
\end{eqnarray}\vspace{1mm}
Equation (\ref{e4}) is the combination $R_{tt}/B+R_{rr}/A$, Eq.
(\ref{e5}) is $R_{rr}/A-R_{\theta\theta}/r^2$, and Eq. (\ref{e6})
is the $rr$-component of the field equation. Finally, Eq.
(\ref{e7}) is from the contraction of Eq. (\ref{e2}). In
principle, for a given $h$ (or$f$) one should be able to solve the
four Eqs. (\ref{e4})-(\ref{e7}) for the four unknowns, $A$, $B$,
$R $, and $f$ (or $h$), as functions of $r$.

\section{Solutions of Equations (4)-(7)}

We are interested in solutions that differ from those of the
classic GR by small amounts. For the classic GR one has $h=1$ and
$A(r)B(r)= 1$. Here, we argue that, if the combination $B'/B+A'/A$
is a well-behaved differential expression, it should have a
solution of the form $A(r)B(r)=g(r)$. Furthermore, $g(r)$ should
differ from 1 only slightly, in order to remain in the vicinity of
GR. There are a host of possibilities. For the sake of argument
let us assume $g(r)=(r/s)^\alpha \approx1+\alpha\ln (r/s)$, where
$\alpha$ is a small dimensionless parameter and $s$ is a length
scale of the system to be identified shortly. Equation (\ref{e4})
splits into \vspace{1mm}\vspace{1mm}
\begin{eqnarray}\label{e8}
&& \frac{B'}{B}+\frac{A'}{A}=\frac{\alpha}{r},
~~~AB=\left(\frac{r}{s}\right)^\alpha,
\\ \label{e9}&&
h''-\frac{1}{2}\frac{\alpha}{r}h'+\frac{\alpha}{r^2}h=0.
\end{eqnarray}
Equation (\ref{e9}) has the solution $h=(r/s)^\beta$,
$\beta=\alpha+O\left(\alpha^2\right)$, and
$1-\frac{1}{2}\alpha+O\left(\alpha^2\right).$ Of these, the
solution $h\approx (r/s)^\alpha$ satisfies the requirement
$h\rightarrow1$ as $\alpha\rightarrow 0$. The second solution is
discarded. Substituting $AB=h=(r/s)^\alpha$ in Eq. (\ref{e5})
gives \vspace{1mm}
\begin{eqnarray} \label{e10}
&&
\frac{1}{A}=\frac{1}{(1-\alpha)}\left[1-\left(\frac{s}{r}\right)^{(1-\alpha/2)}+\lambda
\left(\frac{r}{s}\right)^{2(1-\alpha/2)}\right],\\\vspace{1mm}
\label{e11}&& B=\left(\frac{r}{s}\right)^\alpha \frac{1}{A},
\end{eqnarray}
where $\lambda$ is a constant of integration. Actually there is
another constant of integration multiplying the $(s/r)$ term. We
have, however, absorbed it in the expression for $s$ that we now
define. For $\alpha=0$, Eqs. (\ref{e10}) and (\ref{e11}) are
recognized as the Schwarzschild-deSitter metric. Therefore, $s$ is
identified with the Schwarzschild radius of a central body,
$2GM/c^2$, and $\lambda$ with a dimensionless cosmological
constant. Substitution of Eqs. (\ref{e10}) and (\ref{e11}) into
Eqs. (\ref{e6}) and (\ref{e7}) gives \vspace{1mm}
\begin{eqnarray}
\label{e12} && f=\frac{3}{(1-\alpha)}\frac{1}{r^2}\left[\alpha
\left(\frac{r}{s}\right)^\alpha+(2+\alpha)\lambda
\left(\frac{r}{s}\right)^2 \right],
\\\cr\label{e13} && R=\frac{3}{(1-\alpha)}\frac{1}{r^2} \left[\alpha
+(4-\alpha)\lambda \left(\frac{r}{s}\right)^{(2-\alpha)}\right].
\end{eqnarray}
The Ricci scalar of the Schwarzschild space is zero and that of
the deSitter or the Schwarzschild-deSitter space is constant. For
non zero $\alpha$, however, $R$ is somewhere between these two
extremes. At small distances it increases as $r^{-2}$ and at large
$r$'s it behaves as $s^{-2}(s/r)^\alpha  \approx s^{-2}(1- \alpha
\ln{r/s})$. The spacetime is asymptotically neither flat nor
deSitterian. Cosmologists may find this variable Ricci scalar
relevant to their purpose ( see also Brevik et al, 2004, for a
different modification of the Schwarzschild-deSitter metric).
Likewise, we began with $f$ as a function of $R$ rather than $r$.
Elimination of $r$ between Eqs. (\ref{e12}) and (\ref{e13})
provides one in terms of the other. For $\lambda = 0$, one easily
finds
\begin{eqnarray} \label{e14}
&& f =(3\alpha)^{\alpha/2} s^{-\alpha}R^{(1-\alpha/2)}\approx R
[1-\frac{\alpha}{2}\ln (s^2 R)+\frac{\alpha}{2}\ln(3\alpha)].
\end{eqnarray}
Once more we observe the mild logarithmic correction to the
classic GR.
\section{Applications to galactic environments}

In this section we demonstrate that
\begin{itemize}
\item The logarithmic modification of the Einstein-Hilbert action,
in the weak field regime, results in a logarithmic correction to
the newtonian potential. A test star moving in such a potential
acquires a constant asymptotic speed, $v_\infty=c\sqrt{\alpha/2}$.
\item The asymptotic speed cannot be independent of the central mass.
We resort to the observed rotation curves of spirals to find this
dependence.
\item The high- and low- acceleration limits of the weak-field
regime are the same as those of MOND. A kinship with MOND follows.
\end{itemize}

\subsection{Orbits in the spacetime of Equations (\ref{e10})-(\ref{e13})}

We assume a test star orbiting a central body specified by its
Schwarzschild radius, $2GM/c^2$. We choose the orbit in the plane
$\theta=\pi/2$. The geodesic equations for $r$, $\varphi$, and $t$
are \vspace{1mm}
\begin{eqnarray}\label{e15}
&&
\frac{d^2r}{d\tau^2}+\frac{1}{2}\frac{A'}{A}\left(\frac{dr}{d\tau}\right)^2-\frac{r}{A}\left(\frac{d
\varphi}{d\tau}\right)^2+\frac{1}{2}\frac{B'}{A}\left(\frac{dt}{d\tau}\right)^2=0,
\\\cr\label{e16} &&\left(\frac{d \varphi}{d\tau}\right)^{-1}
\frac{d^2\varphi}{d\tau^2}+\frac{2}{r}\frac{dr}{d\tau}=0,
\\\cr\label{e17} &&
\left(\frac{dt}{d\tau}\right)^{-1}\frac{d^2t}{d\tau^2}+\frac{B'}{B}\frac{dr}{d\tau}=0,
\end{eqnarray}
respectively. Equations (\ref{e16}) and (\ref{e17}) immediately
integrate into
\begin{eqnarray}\label{e18}
&& r^2d\varphi/d\tau=J, ~~\textrm{a constant}, \\\cr \label{e19}
&& dt/d\tau=1/B.
\end{eqnarray}
Substituting the latter into Eq. (\ref{e15}) and assuming a
circular orbit, $dr/d\tau=0$, gives
\begin{eqnarray}\label{e20}
&&
\frac{J^2}{r^3}=\frac{1}{2}\frac{B'}{B^2}=\frac{1}{2}\left(\frac{r}{s}\right)^\alpha\frac{B'}{B^4},
\end{eqnarray}
where we have used Eq. (\ref{e11}) to eliminate A. In galactic
environments what one measures as the circular orbital speed is
\begin{eqnarray}\label{e21}
&& v=\frac{rd\varphi}{\sqrt B dt}=\frac{r}{\sqrt B} \frac{d
\varphi}{d\tau}\frac{d\tau}{d t}=\frac{\sqrt B J}{r}.
\end{eqnarray}
Eliminating $J$ between Eqs. (\ref{e21}) and (\ref{e20}) gives
\begin{eqnarray}\label{e22}
&& v^2=\frac{1}{2} \frac{r B'}{B} = \frac{1}{2}\left[ \alpha -
\frac{rA'}{A} \right].
\end{eqnarray}
Further substitution for $B$ from Eqs. (\ref{e11}) and (\ref{e10})
yields
\begin{eqnarray}\label{e23}
v^2=\frac{1}{2}\alpha + \frac{1}{2}\left(1-\alpha/2\right)
\frac{\left[\left(\frac{s}{r}\right)^{(1-\alpha/2)}+2\lambda
\left(\frac{r}{s}\right)^{2(1-\alpha/2)}\right]}{\left[1-\left(\frac{s}{r}\right)^{(1-\alpha/2)}+\lambda
\left(\frac{r}{s}\right)^{2(1-\alpha/2)}\right]}.
\end{eqnarray}
To put Eq. (\ref{e23}) in a tractable form:

\begin{itemize}
    \item We neglect the $\lambda$ term and substitute
    $s=2GM/c^2$.
    \item We adopt the approximation
$x^{-\alpha}=\exp(-\alpha\ln x)=1-\alpha\ln
x+O\left(\alpha^2\right)$.
    \item The terms containing $s$ are small. We retain
only the first order terms in $s$.
    \item $v$ is measured in units of $c$. We restore it hereafter.
\end{itemize}

With these provisions, Eq. (\ref{e23}) reduces to
\begin{eqnarray}\label{e24}
&& v^2=\frac{1}{2}\alpha c^2
+\frac{GM}{r}\left[1-\frac{1}{2}\alpha\left\{1+\ln\left(\frac{2GM
}{c^2r}\right)\right\}\right].
\end{eqnarray}
A plot of $v^2$ as a function of $r$ has the horizontal asymptote
$\frac{1}{2}\alpha c^2$.
\subsection{Determination of $\alpha$}
The asymptote in Eq. (\ref{e24}) cannot be a universal constant.
It is not possible to imagine that a galaxy and a speck of dust
dictate the same speed for distant passing objects. The parameter
$\alpha$ should depend on the mass of the gravitating body
residing at the origin, because any localized matter will betray
no characteristics other than its mass when sensed from far
distances. To find the mass dependence of $\alpha$ we resort to
observations. From Sanders and Verheijn (1998) and Sanders and
McGaugh (2002), we have compiled a list of 31 spirals for which
total masses, asymptotic orbital speeds, and velocity curves are
reported. The figures in their papers contain the observed
circular speeds and the newtonian ones derived from the observed
mass of the stellar and HI components of the galaxies. We have
selected those objects that a) have a noticeable horizontal
asymptote, b) have fairly reduced newtonian speeds by the time the
flat asymptote is approached, and c) do not possess anomalously
high HI content to hinder estimates of the total mass and the size
of the galaxy. We also made the assumption that the total HI and
stellar mass are distributed spherically symmetrically and mimic a
point mass if observed from far distances. The relevant data along
with $\alpha=2v_\infty^2/c^2$ are reported in the table, and the
figure is a log-log plot of the calculated $\alpha$ versus the
mass. A power law fit to the data gives
\begin{eqnarray}\label{e25}
&&\alpha = (3.07 \pm 0.18)\times 10^{-7} (M/
10^{10}M_\odot)^{0.494}.
\end{eqnarray}\\
It is important to note that Eq. (\ref{e25}) is not a consequence
of the present theory, but rather an empirical relation dictated
by observations and based on the masses and the asymptotic speeds
of a selected list of galaxies reported by Sanders et al. Together
with the popularly accepted rule that the masses and the
luminosities of spirals are linearly related, it leads to a
Tully-Fisher (TF) relation, $Luminosity \propto
{v_{\infty}}^{4.05}$. Observational actualities, however, are
complicated. In a recent paper, Kregel et al. (2005) distinguish
between different TF relations based on the luminosity, disk mass,
maximum disk stellar mass, baryonic mass (meaning stellar+HI
mass), baryonic + bulge mass, etc. The reported exponents range
from $3.23\pm 0.36$ to $4.2\pm 0.23$, depending on the type of
qualification; see also Gurovich et al. (2004). A more elaborate
discussion of the issue falls beyond  the scope of the present
paper.

The main sources of error in Eq. (\ref{e25}), both in the exponent
and in the slope, are a) the estimates of the total masses of the
galaxies, b) the judgment whether what one measures as the
asymptotic speed is indeed the orbital speed at the far outskirts
of the galaxy, c) the popular assumption that the masses and
luminosities of the spirals are linearly related, and finally, d)
our heuristic assumption that the galaxies can be treated as
spherically symmetric objects. In spite of all these
uncertainties, we note that the exponent 0.494 is astonishingly
close to 0.5, the figure that one finds from MOND. We also
demonstrate in the following section that the slope $3.00\times
10^{-7}$ is in very good agreement with the characteristic
acceleration of MOND.

\begin{table}
\caption{ The data in the first four columns are from Sanders et
al. 2002. The last two columns show the empirical relation between
the asymptotic speeds and the masses of the galaxies.}\label{tab1}
\begin{center}
\begin{tabular}{cccccc} \hline
Galaxy&R&M&$v_\infty$&$2(v_\infty/c)^2$&$\alpha_0$\\
&kpc&$10^{10}M_\odot$&km/s&$\times10^{7}$&$\times10^{12}$\\

 \hline\hline
  NGC 5533 & 72.0 & 22.0 & 250 & 13.9 & 3.02 \\
  NGC 3992 & 30.0 & 16.22& 242 & 13.0 & 3.28 \\
  NGC 5907 & 32.0 & 10.8 & 214 & 10.2 & 3.15 \\
  NGC 2998 & 48.0 & 11.3 & 213 & 10.1 & 3.05 \\
  NGC 801  & 60.0 & 12.9 & 218 & 10.6 & 3.00 \\
  NGC 5371 & 40.0 & 12.5 & 208 & 9.61 & 2.76 \\
  NGC 4157 & 26.0 & 5.62 & 185 & 7.61 & 3.24 \\
  NGC 4217 & 14.5 & 4.50 & 178 & 7.04 & 3.35 \\
  NGC 4013 & 27.0 & 4.84 & 177 & 6.96 & 3.19 \\
  NGC 4088 & 18.8 & 4.09 & 173 & 6.65 & 3.32 \\
  NGC 4100 & 19.8 & 4.62 & 164 & 5.98 & 2.81 \\
  NGC 3726 & 28.0 & 3.24 & 162 & 5.83 & 3.26 \\
  NGC 4051 & 10.6 & 3.29 & 159 & 5.62 & 3.12 \\
  NGC 4138 & 13.0 & 3.01 & 147 & 4.82 & 2.80 \\
  NGC 2403 & 19.0 & 1.57 & 134 & 3.99 & 3.19 \\
  UGC 128  & 40.0 & 1.48 & 131 & 3.81 & 3.14 \\
  NGC 3769 & 33.0 & 1.33 & 122 & 3.31 & 2.88 \\
  NGC 6503 & 21.8 & 1.07 & 121 & 3.25 & 3.14 \\
  NGC 4183 & 18.0 & 0.93 & 112 & 2.79 & 2.89 \\
  UGC 6917 & 9.0  & 0.74 & 110 & 2.69 & 3.12 \\
  UGC 6930 & 14.5 & 0.73 & 110 & 2.69 & 3.14 \\
  M   33   & 9.0  & 0.61 & 107 & 2.54 & 3.24 \\
  UGC 6983 & 13.8 & 0.86 & 107 & 2.54 & 2.74 \\
  NGC 7793 & 6.8  & 0.51 & 100 & 2.22 & 3.10 \\
  NGC 300  & 12.4 & 0.35 & 90  & 1.80 & 3.02 \\
  NGC 5585 & 12.0 & 0.37 & 90  & 1.80 & 2.94 \\
  NGC 6399 & 6.8  & 0.28 & 88  & 1.72 & 3.22  \\
  NGC 55   & 10.0 & 0.23 & 86  & 1.64 & 3.39 \\
  UGC 6667 & 6.8  & 0.33 & 86  & 1.64 & 2.83 \\
  UGC 6923 & 4.5  & 0.24 & 81  & 1.46 & 2.95 \\
  UGC 6818 & 6.0  & 0.14 & 73  & 1.18 & 3.12 \\
 \hline
  \end{tabular}
  \end{center}
R: radius of the galaxy (kpc); M: stellar + HI mass
($10^{10}M_\odot$); $v_{\infty}$: asymptotic speed  (km/sec);
$\alpha_0$: $2(v_\infty/c)^2 M^{-0.494}$.

 \end{table}
 \begin{center}
\begin{figure}[h]
\includegraphics{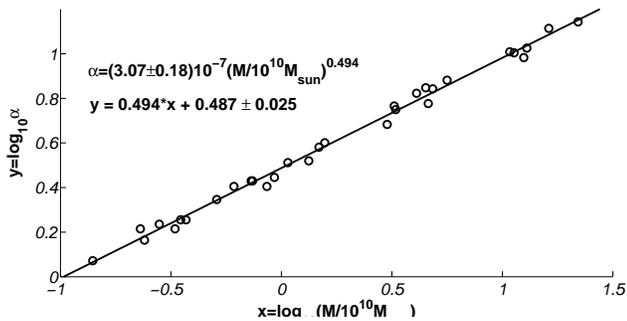} \vspace{4.3cm} \caption {A log-log plot of $\alpha$
versus M. The equation for the power law fit is shown in the
legend. }\label{f1}
   \end{figure}
     \end{center}

\subsection{ Kinship with MOND}

We recall that in the weak-field approximation, the newtonian
dynamics is derived  from the Einsteinian one by writing the
metric coefficient $B=\left(1+2\phi/c^2\right),\phi=GM/r$ and by
expanding all relevant functions and equations up to the first
order in $\phi/c^2$. In a similar way one may find our modified
newtonian dynamics from the presently modified GR by expanding
$B(r)$ of Eq. (\ref{e11}) up the first order in $\alpha$ and
$s/r$. Thus
\begin{eqnarray}\label{e26}
 &&B(r) =1+\alpha+\alpha\ln(r/s)-s/r=1+2\phi(r)/c^2,
\end{eqnarray}
where the second equality defines $\phi(r)$. Let us write Eq.
($\ref {e25}$) (with slight tolerance) as $\alpha=\alpha_0
(GM/GM_\odot)^{1/2}$ and find the gravitational
acceleration\\
\begin{eqnarray}\label{e27}
~~~g \al~= \al~ \left|d\phi/dr\right|=(a_0 g_n)^{1/2}+g_n \\
\al~ = \al~g_n ~~{\rm for}~~ g_n\gg a_0 \cr \al~ = \al~ (a_0
g_n)^{1/2}~~ {\rm for} ~~ a_0\gg g_n \rightarrow 0, \nonumber
\end{eqnarray}
where we have denoted
\begin{eqnarray}\label{e28}
&&a_0  =  \alpha_0 ^2c^4/4GM_\odot ~~{\rm and}~~ g_n=GM/r^2.
\end{eqnarray}
The limiting behaviors of $g$ are the same as those of MOND. One
may, therefore, comfortably identify $a_0$ as MOND's
characteristic acceleration and calculate $\alpha_0$ anew from Eq
(\ref{e28}). For $a_0=1.2\times10^{-8}$cm/sec${}^2$, one finds
\begin{eqnarray}\label{e29}
 &&\alpha = 2.8\times 10^{-12} \left(M/M_\odot\right)^{1/2}.
\end{eqnarray}
\indent It is gratifying how close this value of $\alpha$ is to
the one in Eq. ($\ref {e25}$) and how similar the  low and high
acceleration limits of MOND and the present formalism are, in
spite of their totally different and independent starting points.
It should also be noted that there is no counterpart to the
interpolating function of MOND here.
\section{Concluding remarks}
We have developed an $f(R)\propto R^{1-\alpha/2}$ gravitation that
is essentially a logarithmic modification of the Einstein-Hilbert
action. In spherically-symmetric static situations, the theory
allows a modified Schwarzschild-deSitter metric. This metric in
the limit of weak fields gives a logarithmic correction to the
newtonian potential. From the observed asymptotic speeds and
masses of spirals we learn that the correction is proportional to
almost the square root of the mass of the central body. Flat
rotation curves, the Tully-Fisher relation (admittedly with some
reservations), and a version of MOND emerge as natural
consequences of the theory.

Actions are ordinarily form invariant under the changes in
sources. Mass dependence of $\alpha$ destroys this feature and any
claim for the action-based theory should be qualified with such
reservation in mind. This, however, should not be surprising, for
it is understood that all alternative gravitations, one way or
another, go beyond the classic GR. One should not be surprised if
some of the commonly accepted notions require re-thinking and
generalizations.

Since the appearance of an earlier version of this paper in arXiv,
Mendoza et al.(2006) have investigated the gravitational waves and
lensing effects in the proposed spacetime. They find the
following: a) in any $f(R)=R^n$ gravitation, gravitational waves
travel with the speed of light in a vacuum, and b) in the present
spacetime, there is a lensing in addition to what one finds in the
classical GR. Their ratio of the additional deflection angle of a
light ray, $\delta\beta$, to that in GR, $\beta_{GR}$, can be
reduced to
\begin{eqnarray}\label{e31}
&& \delta\beta/\beta_{GR} =\frac{1}{2}\alpha \ln {(r_m/s-1)},
\end{eqnarray}
where $r_m$ is the impact parameter of the impinging light. The
proportionality of $\delta\beta$ to $\alpha$ is expected, because
the proposed metric is in the neighborhood of GR. Its increase
with increasing $r_m$ also should not be surprising, since the
theory is designed to highlight unexpected features at far rather
than nearby distances.

Soussa et al. (2004) maintain that ``no purely metric-based,
relativistic formulation of MOND whose energy functional is stable
can be consistent with the observed amount of gravitational
lensing from galaxies". For at least two reasons, this no-go
theorem does not apply to what we have highlighted above  as the
kinship with MOND:

\noindent a) Apart from their common low and high acceleration
regimes, the two theories are fundamentally different. The
gravitational acceleration in the weak field limit of the present
theory is the newtonian one to which a small $1/r$ correction is
added. That of MOND, on the other hand, is a highly nonlinear
function of the newtonian acceleration through an arbitrary
interpolating function.

\noindent b) More important, however, is one of the authors'
assumptions that ``the gravitational force is carried by the
metric, and its source is the usual stress tensor". This is not
the case in the present theory. Although we have only worked out
the vacuum solution for a point source, the mass dependence of the
exponent $\alpha$ in Eqs. (\ref{e10}) and (\ref{e11}) makes the
theory different from what the assumption requires.

There are two practices for obtaining the field equations of
$f(R)$ gravity, the metric approach, where $g_{\mu\nu}$'s are
considered as dynamical variables, and that of Palatini, where the
metric and the affine connections are treated as such (see Magnano
1995 for a review). Unless $f(R)$ is linear in $R$, the resulting
field equations are not identical (see Ferraris et al. 1994). The
metric approach is often avoided for leading to fourth-order
differential equations. It is also believed to have instabilities
in the weak field approximations (see e. g., Sotiriou 2005 and
also Amarzguioui et al. 2005). In the present paper we do not
initially specify $f(R)$. Instead, at some intermediate stage in
the analysis we adopt an ansatz  for $df(R)/dR$ as a function of
$r$ and work from there to obtain the metric, the Ricci scalar,
and eventually $f(R)$. This enables us to avoid the fourth-order
equations. This trick should work in other contexts, such as
cosmological ones.

The theory presented here is preliminary. Further investigations
are needed from both formal and astrophysical points of view. The
author's list of priorities include the following:
\begin{itemize}
\item Stability of the metric of Eqs. (\ref{e10}) \&(\ref{e11}).
The approach should be to impose a small perturbation $\delta
g_{\mu \nu}$ on the metric, linearize the field Eq. (\ref{e2}),
and ask for the condition of stability of the metric. Such a
condition, if it exists at all, might throw some light on the mass
dependence of $ \alpha $, the empirical relation of Eq.
(\ref{e29}). Managing the linear problems is straightforward.
Here, however, the bookkeeping is extensive and laborious.
\item Extension of the theory, at least in the weak field regime, to
many body systems and to cases with a continuous distribution of
matter, in order to obtain the metric inside the matter.
\item Developing the theory beyond the first order in $\alpha$
\item Solar system tests of the theory.
\item Possible cosmological implications of the theory.
\end{itemize}


\noindent \textbf{Acknowledgement}: The author wishes to thank
Bahram  Mashhoun and Naresh Dadhich for comments and helpful
suggestions. Reza Saffari has pointed out a typographical error in
the earlier version of the paper, Eq. (15). The error had
propagated making the sign of $\alpha$ appear positive in the
factor multiplying the term $GM/r$ in Eq. (24). This is corrected
here.
{}


\end{document}